\documentclass[aps,pra,onecolumn]{revtex4-1}

\usepackage[normalem]{ulem}
\usepackage{amsmath,amssymb}
\usepackage{multirow}
\usepackage{xcolor,soul}
\usepackage{srcltx}
\usepackage{hyperref,graphicx}

\begin{document}

\title{Floquet bound states in the continuum}

\author{Stefano Longhi}
\email{Corresponding author email: stefano.longhi@polimi.it}

\author{Giuseppe Della Valle}

\vspace{3cm}
\affiliation{Dipartimento di Fisica- Politecnico di Milano and Istituto di Fotonica e Nanotecnologie - Consiglio Nazionale delle Ricerche \ \\ Piazza Leonardo da Vinci, 32, I-20133 Milano, Italy}
\begin{abstract} {\bf Quantum mechanics predicts that certain stationary potentials can sustain bound states with an energy buried in the continuous spectrum of scattered states, the so-called bound states in the continuum (BIC). Originally regarded as mathematical curiosities, BIC have found an increasing interest in recent years, particularly  in quantum and classical transport  of matter and optical waves in mesoscopic and photonic systems where the underlying potential can be judiciously tailored. Most of our knowledge of BIC is so far restricted to static potentials. Here we introduce a new kind of BIC, referred to as  Floquet BIC, which corresponds to a normalizable Floquet state of a time-periodic Hamiltonian with a quasienergy embedded into the spectrum of Floquet scattered states. We discuss the appearance of Floquet BIC states in a tight-binding lattice model driven by an ac field in the proximity of the dynamic localization regime.}
\end{abstract}
\maketitle
At the birth of quantum mechanics, von Neumann and Wigner \cite{NW} suggested rather surprisingly that certain spatially oscillating attractive potentials can sustain normalizable states at a positive energy, i.e. embedded into the spectrum of scattered states.  Because of their unusual geometry, such potentials were earlier regarded as mathematical curiosities with low physical relevance, and for many years BIC did not attract the interest of the scientific community. Subsequent theoretical and experimental studies showed that BIC can be found in a wide range of different quantum and classical systems,  including atomic and molecular  systems \cite{a1,a2,a3,a4}, semiconductor and mesoscopic structures \cite{m0,m1,m1bis,m2,m3,m4,m5}, graphene \cite{gr}, quantum Hall insulators \cite{NC},  optical structures \cite{o1,o2,o3,o4}, and Hubbard models \cite{H1,H2}. Generally, BIC are fragile states, which  decay into resonance states by small perturbations. However, they can play an important role in quantum transport on the nanoscale with application to nanoelectronics \cite{orell} and spintronics \cite{spin},  or in the design of photonic structures for enhancement 
of nonlinear phenomena with application to biosensing and
impurity detection \cite{m1}. In some cases BIC possess a certain degree of robustness against hybridization into the continuum \cite{NC,H1,molina}, and can survive under  harmonic modulation \cite{orell}. 
As the existence of BIC states in static potentials is well established, the possibility to observe BIC states in time-periodic Hamiltonians has not received great attention to date.  Time periodic Hamiltonians are found in a wide range of different physical fields and describe important physical phenomena,  for example driven quantum tunneling, scattering from oscillating potentials and quantum transport on the nanoscale \cite{rev1,rev2}. Application of ac fields has become a very promising tool to
engineer quantum systems, for example to effectively simulate properties of undriven
systems in higher dimensions \cite{Platero}, to control topological states of matter and to induce topological insulators \cite{top1,top2}.

Here we introduce a new kind of BIC, referred to as  Floquet BIC, which correspond to breathing normalizable Floquet states of a time-periodic Hamiltonian with a quasienergy embedded into the spectrum of Floquet scattered (non-normalizable) states. We discuss the existence of such BIC states for a quantum particle in a tight-binding lattice model driven by an ac field. Floquet BIC are found under certain driving conditions in the neighborhood of the dynamic localization regime. In the high-frequency limit, such states can be explained as a result of selective destruction of tunneling. 

\section*{Results}
\noindent\textbf{Driven lattice model.}  As a model system, we consider the coherent hopping dynamics of a quantum particle on a one-dimensional tight-binding lattice driven by an external sinusoidal field with inhomogeneous hopping rates, which is described by the Hamiltonian (with $\hbar=1$)
\begin{equation}
\hat{H}= \sum_n \kappa_n \left\{ | n \rangle \langle n+1| +| n+1 \rangle \langle n | \right\}+F(t) a \sum_n n |n \rangle \langle n |
\end{equation}
where $|n \rangle$ is the Wannier state localized at lattice site $n$ ($n=0, \pm 1, \pm 2,...$), $\kappa_n$ is the hopping rate between sites $n$ and $(n+1)$, $a$ is the lattice period, and $F(t)=F_0 \cos (\omega t)$ is the external sinusoidal force of period $T= 2 \pi / \omega$. This model has been investigated in different physical contexts. It describes, for example, coherent transport of ultracold atoms  in periodically-shaken optical lattices \cite{Arim}, coherent electronic transport in irradiated semiconductor superlattices \cite{Holthaus}, and light propagation in arrays  of periodically-curved optical waveguides \cite{Marangoni,PRep}.
We assume that the lattice is asymptotically homogeneous, i.e. that $\kappa_n \rightarrow \kappa$ as $ n \rightarrow \pm \infty$, and that in the absence of the driving force, i.e. for $F(t)=0$, the static Hamiltonian $\hat{H}$ has a purely continuous spectrum, i.e. that hopping inhomogeneities do not introduce bound states, neither outside nor inside the tight-binding energy band $(-2 \kappa, 2 \kappa)$. For example, such a condition is satisfied by assuming $\kappa_n=\kappa$ for $ n \neq -1,0$ and $ \kappa_{-1}=\kappa_0=\rho < \kappa$, see Fig.1(a). In the presence of the periodic forcing, the eigenstates $| \psi (t) \rangle$ of the Hamiltonian are Floquet states of the form $|\psi (t) \rangle=| u (t) \rangle \exp(-i \epsilon t / \hbar)$, where $|u(t+T) \rangle=|u(t) \rangle$ and $\epsilon$ is the quasi-energy, which is assumed to vary in the interval $(-\omega/2, \omega/2)$. A Floquet BIC state can be defined as a normalizable Floquet state of $\hat{H}$ with a quasi-energy buried in the continuous quasi-energy spectrum of scattered states of the lattice. For a homogeneous lattice ($\kappa_n=\kappa$), the problem is integrable \cite{Dunlap} and the quasi-energy spectrum turns out to be purely continuous and defined by the dispersion relation \cite{Holthaus} $\epsilon(p)=2 \kappa J_0(\Gamma) \cos(p)$, where $-\pi \leq p < \pi$, $\Gamma= aF_0 /   \omega$ is the normalized forcing parameter, and $J_0$ is the Bessel function of first kind and zero order.  The two linearly-independent Floquet scattered states with quasi-energy $\epsilon( \pm p)$ in the ac-driven homogeneous lattice are given by the backward and forward propagating plane waves
\begin{equation}
|\psi_{ \pm p}(t) \rangle = \sum_n \exp [  \pm ip n-i \Theta(\pm p,t) ] | n \rangle,
\end{equation}
 where 
$\Theta(p,t)=\Phi(t)+2 \kappa \int_0^t dt' \cos [p-\Phi(t')]$ and $\Phi(t)=\Gamma \sin (\omega t)$.
As is well known, the quasi-energy spectrum shrinks and collapse as $\Gamma \rightarrow 2.405$, at which $J_0(\Gamma)=0$ and dynamic localization (DL)  is attained \cite{Dunlap}. Hence, in a homogeneous lattice the role of the external force is to re-normalize the bandwidth of the undriven lattice band, however the spectrum remains purely continuous. Lattice defects or boundary effects make DL imperfect and bound states outside the continuum (BOC) can be  induced by the ac field, as discussed in \cite{B1,B2,B3}. Here we show that, besides BOC states, under certain driving conditions Floquet BIC states can appear as well.

\vspace{0.3cm}

\noindent\textbf{Floquet bound states in the continuum.} 
To this aim, we considered the lattice model of Fig.1(a) assuming $\rho / \kappa =0.7$, and numerically computed the quasi-energy spectrum and associated Floquet eigenstates for a few values of the ratio $\kappa / \omega$ and for a normalized driving amplitue $\Gamma$ spanning the interval $(2,2.8)$, near the DL point $\Gamma=2.405$. The lattice comprises $N=201$ sites. To avoid lattice truncation effects, periodic boundary conditions have been assumed in the numerical analysis (see Methods). The degree of localization of the Floquet states  $|u(t) \rangle$ is described by the participation ratio $R(t)$, which is defined as $R(t)=\langle u(t) |u(t) \rangle^2 / \langle u^2(t) |u^2(t) \rangle$.  Note that $R(t)$ is periodic with period $T$, with $R \sim 1$ for localized modes while $R \sim N$ for extended states.  For the undriven lattice, all the modes are extended (with $R \sim 134$), and there are no signature of resonances; see Fig.1(b). The scenario is fully modified when the driving field is switched on. Figure 2 shows the numerically-computed quasi-energy spectrum and participation ratio $R(t)$ (in a log scale) at $t=0$ of the corresponding $N$ Floquet modes versus the normalized forcing $\Gamma$ for the three values $\kappa / \omega=2$, $\kappa / \omega=1$ and $\kappa / \omega=0.3$. The eigenmode number of Floquet states is ordered for increasing values of the quasi energies. An inspection of the quasi energy diagrams shows that, in the neighborhood of the DL regime $\Gamma=2.405$,  Floquet BOC emerge in pairs, above and below the band of scattered states, which are clearly visible as isolated dispersion curves that detach from the continuous band of scattered states. In the participation ratio diagrams, the BOC modes correspond to the dark stripes at the upper and lower boundaries of the domain. The number of field-induced BOC states typically increases as the ratio $\kappa/ \omega$ increases, according to previous studies \cite{B2}. As $ \Gamma$ is pushed far from the DL condition, the dispersion curves of the BOC states go inside the band of scattered states, and resonance states, i.e. unbounded states which retain a certain degree of localization, are clearly visible in the participation ratio diagrams as darker lines internal to the domain. Examples of BOC and resonance states are shown in Figs.3(a) and (b), respectively. 

As BOC states are bound states with exponential decay tails as $ n \rightarrow \pm \infty$, resonance states show non-decaying (oscillatory) tails, which far from the defect  region (i.e. as $n \rightarrow \pm \infty$) are asymptotically given by a superposition of the forward and backward propagating plane waves $| \psi_{\pm p}(t) \rangle$ of the ac-driven uniform lattice, defined by Eq.(2). The wave number $p$ of the oscillating tails is related to the quasi energy $\epsilon$ of the resonance state via the relation $\epsilon=-2 \kappa J_0(\Gamma) \cos p$. Obviously for a resonance state, like for any unbounded state,  $R$ diverges as $N \rightarrow \infty$. A degree of localization of a resonance state, which is independent of the number $N$ of lattice sites, can be defined by the ratio $1/D=h_1/h_2$ between the absolute maximum, at the reference time $t=0$, of the site occupation probabilities $| \langle \psi(t=0)|n \rangle|^2$, i.e. $h_1={\rm max}_{n}  | \langle \psi(t=0)|n \rangle|^2$, and the maximum of the same quantity  in the tails of the mode, i.e. $h_2={\rm max}_{|n| \gg 1}  | \langle \psi(t=0)|n \rangle|^2$, see Fig.3(b). A BIC state corresponds to the limit of a resonance state with $D \rightarrow 0$, i.e. to the absence of excitation of lattice sites far from the defective region. Such a circumstance is unlikely, and indeed for a general driving force $\Gamma$ BIC can not be found. However, extended numerical simulations showed that, at some special values of $\Gamma$, resonances with extremely small values of $D$ ($D< 10^{-6}$) arise, which can be regarded as BIC states within numerical accuracy. As a general rule, like for BOC states the number of BIC states increases as the ratio $\kappa / \omega$ is increased, i.e. in the low-frequency regime. BIC appear in pairs, at quasi energies $ \pm \epsilon$, or as isolated states with quasi-energy $\epsilon=0$. In the high-frequency regime, the number of BIC states is always 2, and they appear as isolated states with  quasi energy $\epsilon=0$. They are found at two values of the driving field $\Gamma$, one below and the other above the DL condition. The physical origin of such two BIC states will be discussed below and related to the phenomenon of selective destruction of tunneling. As the modulation frequency is decreased, additional pairs of BIC states are generated. Specifically, the number of BIC states found as the forcing $\Gamma$ is spanned over the range $(2,2.8)$ is two, four and eight for $\kappa / \omega=0.3$, $\kappa / \omega=1$ and $\kappa / \omega=2$, respectively.
The locations  of such BIC states are indicated by open circles in the diagrams of Fig.2. In Fig.3(c), (d) and (e) we show a few examples of the occupation probability distributions $| \langle \psi(t) | n \rangle |^2$ of BIC states at $t=0$ (upper panels) and over one oscillation cycle (lower panels), in the low [Fig.3(c)], intermediate [Fig.3(d)] and high-frequency [Fig.3(e)] regimes, corresponding to the states labelled by A,B and C in Fig.2.

\vspace{0.3cm}

\noindent\textbf{Selective destruction of tunneling.} 
For static Hamiltonians, several mechanisms of formation of BIC states have been suggested, including symmetry constraints of the system
\cite{o4,cazzo1} and destructive interference of resonance modes coupled to a common continuum \cite{a3}. The latter mechanism is generally at the origin of BIC modes in stationary tight-binding lattice models \cite{m3,m4,gr,o1}. What is the physical mechanism leading to Floquet BIC states observed in the ac-driven lattice? Before answering to this question, let us first notice that the undriven lattice model (1) does not sustain any BIC, nor BOC modes or long-lived resonance states. Hence, the mechanism of field-induced  BIC formation is a  genuine non-perturbative effect of the time-periodic Hamiltonian, and can not be mapped into the topology of the undriven lattice  (like, for example, in the model discussed in Ref.\cite{orell}). The effect of the driving field is to extend the effective dimensionality of the lattice because of the different Floquet channels introduced by the ac field \cite{Platero}. Precisely, Floquet BIC states of the ac-driven 1D lattice model (1) can be mapped into BIC states of a static 2D lattice model with non-neighboring hopping rates and an effective electric field applied along the Floquet channels, see Ref.\cite{Platero}. In such an effective static 2D lattice model, the appearance of BIC states can be rather generally explained as a result of destructive interference among different tunneling paths that enables to localize the excitation in a finite number of lattice sites. Regrettably, the complexity  of the hopping rates in the effective 2D lattice does not permit to provide a simple description of BIC formation. However, in the high-frequency limit the ac-driven lattice model can be mapped into an effective static lattice model with the same dimensionality of the undriven lattice \cite{LonghiPRB08}, and 
the mechanism of BIC formation can be explained as a result of {\it selective destruction of tunneling} that occurs at special driving conditions \cite{LonghiPRB08,Villas}. In fact, in the high-frequency limit $\omega / \kappa \gg 1$, the ac-driven lattice of Fig.1(a) is equivalent to the static 1D lattice of Fig.1(c) with hopping rates $\kappa_e$, $\alpha$ and $\beta$  given by (see Methods)
\begin{eqnarray}
\kappa_e (\Gamma) & = & \kappa J_0(\Gamma) \nonumber \\
\alpha (\Gamma) & = & \kappa J_0(\Gamma)- \kappa Q (\Gamma) (\rho^2-\kappa^2) \\
 \beta (\Gamma) &= &  \rho  J_0(\Gamma)+\rho Q (\Gamma) (\rho^2-\kappa^2) \nonumber
\end{eqnarray}
where the function $Q (\Gamma)$ is defined by Eq.(5) given in the Methods. When $\Gamma$ varies in the interval $(2,2.8)$, the function $Q (\Gamma)$ turns out to be negative and bounded from below, namely $Q(\Gamma)>-0.305$. On the other hand, $J_0(\Gamma)$ changes sign as the DL point $\Gamma=0$ is crossed. For $\rho < \kappa$, it follows that there exist two values $\Gamma_1$ and $\Gamma_2$ of $\Gamma$, the former below and the latter above the DL point $\Gamma=2.405$, at which $\alpha(\Gamma_1)=0$ and $\beta (\Gamma_2)=0$. The appearance of the two BIC states found in Fig.2(c) can be then explained on the basis of the effective lattice model of Fig.1(c)  as a result of selective destruction of tunneling that occurs at $\Gamma=\Gamma_1$  or at $\Gamma=\Gamma_2$. In fact, at $\Gamma=\Gamma_1$, one has $\alpha=0$, so that the three sites $|-1 \rangle$, $|0 \rangle$ and $|1 \rangle$ are effectively decoupled from the other lattice sites. The three eigenstates of the trimer correspond to two BOC states and to one BIC state, namely $|\psi \rangle=(1/ \sqrt{2}) ( |-1 \rangle- |1 \rangle)$, which is precisely the BIC state shown in Fig.3(e) with quasi energy $\epsilon=0$. For $\Gamma=\Gamma_2$, one has $\beta=0$, i.e. the site $|0 \rangle$ is effectively decoupled from the other lattice sites. In this case the BIC merely corresponds to the excitation localized in the $|0 \rangle$ lattice site, whereas four BOC states are sustained at the boundary of the two semi-infinite lattices decoupled from the site $|0 \rangle$. Selective destruction of tunneling provides a simple explanation of the appearance of BIC states in the high-frequency regime, however in the intermediate or low frequency regimes BIC Floquet states {\it do not arise} from coherent suppression of tunneling. In fact, selective destruction of tunneling  implies that the lattice is effectively broken, and a wave packet propagating along it is fully reflected at the lattice site where the effective hopping rate vanishes. Indeed, this is what we observed in the high-frequency regime, in agreement with the effective lattice model of Fig.1(c). In Figure 4(a) we show the numerically-computed evolution of an initial Gaussian wave packet that propagates along the ac-driven lattice with momentum $p= \pi /2$ and mean quasi-energy $\epsilon=0$, for $\kappa / \omega= 0.3$ and $\Gamma=2.308$. As expected, the wave packet is fully reflected when it reaches the site $|-2 \rangle$, because in the effective lattice model the hopping rate $\alpha$ vanishes and the wave packet is forbidden to cross the lattice. Figure 4(b) shows the numerically-computed evolution of the same Gaussian wave packet, but in the ac-driven lattice corresponding to $\kappa / \omega= 2$ and $\Gamma=1.994$, i.e. in the low-frequency regime. At this forcing amplitude the lattice sustains two BIC modes, see Fig.2(a) and 3(c). However, as opposed to the case of Fig.4(a), the wave packet can propagate along the lattice, and reflection at the defective region is rather small. This shows that the ac field does not suppress tunneling among adjacent sites in the lattice, like in the case of Fig.4(a). Hence in the low-frequency regime the appearance of Floquet BIC states is not trivially related to selective destruction of tunneling, and the ac-driven lattice behaves effectively as a 2D lattice with non-trivial states \cite{Platero}.

\section*{Discussion}
Transport and scattering dynamics in periodically-driven systems is of fundamental importance in different areas of physics. 
In time periodic Hamiltonians, the energy of the particle is not conserved owing
to the absorption or emission of quanta from the driving field. This can disclose a wide range of important physical phenomena, such as  chaotic
scattering and chaos-assisted tunneling, coherent destruction of
tunneling, Fano resonances, and field-induced topological insulators, just to mention a few.
Here we have introduced a new phenomenon peculiar to a time-period quantum system, namely the appearance of Floquet bound states in the continuum. Floquet BIC are bound (normalized) Floquet states of a time-periodic Hamiltonian with a quasi energy buried in the quasi energy spectrum of scattered (non-normalized) Floquet states. 
The concept of BIC was proposed in the early days of quantum mechanics by Neumann and Wigner in a seminal paper \cite{NW}, and BIC have been found in a wide range of different physical systems, including atomic and molecular systems, semiconductor and mesoscopic structures, quantum dot chains,  and photonic systems. However, until now the concept of BIC was limited to time-independent Hamiltonians. Here we have considered a widely studied time-periodic Hamiltonian on a lattice, which describes coherent hopping dynamics of a quantum particle in an ac-driven one-dimensional tight-binding lattice. Our main result is that under special driving conditions Floquet BIC can be found in this very simple model, provided that the periodicity of the lattice is broken by the introuduction of a defect.  In the high-frequency regime, the mechanism responsible for the appearance of Floquet BIC is selective destruction of tunneling. In this case the BIC confines the excitation in a few lattice sites, which are effectively decoupled from the other lattice sites thus interrupting the transport along the lattice. However, in the low-frequency regime the BIC states are transparent, and the mechanism of their formation is a nontrivial one and should be traced back to the higher dimensionality of the lattice induced by the ac field. As compared to BIC in a static lattice, the appearance or disappearance of Floquet BIC in a periodically-driven lattice can be externally controlled by the ac field. It is envisaged that the idea of Floquet BIC states disclosed in this work can be extended to a wider kinds of time-periodic Hamiltonians, thus motivating further theoretical and experimental studies. On the experimental side,  Floquet BIC states should be observable in quantum or classical simulators of the the tight-binding Hamiltonian (1), such as ultracold atoms trapped in a periodically-shaken (accelerated) optical lattice or light transport in periodically-curved waveguide arrays. \par

\section*{Methods}

{\bf Numerical method.} Bound Floquet states and corresponding quasi energies of the lattice Hamiltonian (1) have been computed by expanding the vector state of the system $| \psi(t) \rangle$ on the Wannier basis $|n \rangle$ as 
$|\psi(t) \rangle=\sum_{n=1}^{N} c_n(t) \exp [ in \Phi(t)]$ and by numerically solving the following coupled equations for the occupation amplitudes $c_n$
\begin{equation}
i \frac{dc_n}{dt}=\kappa{_n} c_{n+1} \exp[i \Phi(t)]+\kappa_{n-1} c_{n-1} \exp[-i \Phi(t)].
\end{equation}
Equations (4) are solved over one oscillation cycle $(0,T)$ with initial conditions $c_{n}(0)=\delta_{n,m}$, $m=1,2,3,....,N$. To avoid lattice truncation effects, periodic boundary conditions have been used by assuming $c_{0}(t)=c_{N}(t)$ and $c_{N+1}(t)=c_{1}(t)$ in Eq.(4). The Floquet quasi energies $\epsilon$ and corresponding Floquet eigenmodes are computed from the eigenvalues and eigenvectors of the $N \times N$ propagator matrix $S_{n,m}$, where $S_{n,m}$ is the value $c_n(T)$ obtained after solving Eq.(4) with the initial condition $c_{l}(0)=\delta_{l,m}$. The degree of localization of a Floquet eigenstate is measured by its participation ratio $R(t)$, which is defined by $R(t)= (\sum_n |c_n(t)|^2)^2 / \sum_n |c_n(t)|^4$. For unbound states, $R$ diverges as $N \rightarrow \infty$, whereas it remains bound for a localized state. A resonance state is defined as a Floquet eigenstate which is unbounded and shows oscillating (non-decaying) tails, however excitation is preferentially localized at the defective region near the $n=0$ lattice site. A degree of localization of a resonance state, which does not diverges as the number $N$ of lattice sites increases, is the ratio  $1/D=h_1/h_2$, where $h_1$ and $h_2$ are defined in Fig.3(b). A BIC Floquet state corresponds to a resonance state with $D=0$. In our numerical simulations, Floquet BIC states were identified with accuracy $D<10^{-6}$.  

\vspace{0.1cm}
{\bf Effective stationary lattice in the high-frequency regime.} In the high-frequency limit $\epsilon = \kappa / \omega \ll 1$, the ac-driven 1D lattice model (1) can be mapped into an equivalent 1D static lattice with modified hopping rates. This can be formally shown by a multiple-time-scale asymptotic analysis of Eq.(4), after introduction of the normalized time $\tau= \omega t$ \cite{LonghiPRB08}. We look for a solution to Eq.(4) in the form of a power series $c_n(\tau)=A_n(\tau)+ \epsilon c_n^{(1)}(\tau)+ \epsilon^2 c_n^{(2)}(\tau)+...$, where the amplitudes $A_n(\tau)$ vary on slow time scales. Assuming a driving force $\Gamma$ close to DL point, such that $J_0(\Gamma) \sim \epsilon^2$, the evolution of the amplitudes $A_n$ occurs on the slow time scale $\sim \epsilon^3 \tau$ and is found by pushing the asymptotic analysis up to the order $\sim \epsilon^3$ \cite{LonghiPRB08}. 
The resulting evolution equations for the slowly-varying amplitudes $A_n$ describe an effective stationary 1D lattice with Hamiltonian  
\begin{equation}
\hat{H}_{eff}= \sum_n \Delta_n \left\{ | n \rangle \langle n+1| +| n+1 \rangle \langle n | \right\}
\end{equation}
where the effective hopping rate $\Delta_n$ between lattice sites $|n \rangle$ and $|n+1 \rangle$ is given by
\begin{equation}
\Delta_n=\kappa_n J_0(\Gamma)- Q (\Gamma) \left[ \kappa_n \kappa_{n+1}^2-2 \kappa_n^3+ \kappa_n \kappa_{n-1}^2 \right]
\end{equation}
and where we have set 
\begin{equation}
Q (\Gamma) \equiv -(1 / \omega^2) \sum_{l,j \neq 0} J_l(\Gamma) J_j(\Gamma) J_{j-l}(\Gamma) / lj.
\end{equation}
 For the lattice of Fig.1(a), i.e. for $\kappa_n=\kappa$ for $n \neq 0,-1$ and $\kappa_0=\kappa_{-1}= \rho$, the effective static lattice, described by the Hamiltonian (5), is the one depicted in Fig.1(c) with effective hopping rates $\kappa_e$, $\alpha$ and $\beta$ defined by Eqs.(3).

\bibliography{MEP_BIB}

{\small 
\acknowledgements{
This work was partially supported by the Fondazione Cariplo (Grant
No. 2011-0338).}

\section*{Author contributions}
SL developed the theory and GDV made the numerical simulations. Both authors discussed the results and participated in the manuscript preparation.

%\section*{Additional information}
%Supplementary information is available in the online version of the paper.

\section*{Competing financial interests}
The authors declare no competing financial interests.

\newpage

\section*{Figure Captions}
\vspace{1cm}
\noindent
{\bf Fig.1} (a) Schematic of an ac-driven tight-binding lattice with inhomogeneous hopping rate $\rho< \kappa$ between lattice sites $|0 \rangle$ and $| \pm 1 \rangle$. (b)
Participation ratio $R$ of the eigenstates of the undriven lattice comprising $N=201$ sites for $\rho / \kappa=0.7$. The mode number on the horizontal axis is ordered for increasing values of the energy $\mathcal{E}$, from $\mathcal{E}=-2 \kappa$ (lower axis limit) to $\mathcal{E}=2 \kappa$ (upper axis limit). (c) Effective static lattice model in the high-frequency regime. The effective hopping rates $\kappa_e$, $\alpha$ and $\beta$ are given in the text by Eq.(3).\\
\\
{\bf Fig.2} Numerically-computed quasi-energy spectrum versus the normalized forcing amplitude $\Gamma=F_0a/ \omega$ (upper panels) and corresponding maps of the participation ratio $R(t)$ at time $t=0$ of the $N=201$ Floquet eigenstates (lower panels) in the ac-driven lattice of Fig.1(a)  for (a) $\kappa / \omega=2$, (b) $\kappa / \omega=1$, and (c) $\kappa / \omega=0.3$. The mode number on the vertical axis in the lower panels is ordered for increasing values of the quasi energy $\mathcal{E}$. In the upper panels, the dispersion curves that detach from the continuous band correspond to Floquet BOC modes. The circles in the lower panels indicate Floquet BIC states. The BIC modes are found at the forcing amplitudes $\Gamma=1.9940, 2.3121475, 2.35972, 2.49582, 2.6815$ in (a), $\Gamma=2.2904, 2.356951, 2.5195$ in (b), and $\Gamma=2.3800, 2.42875$ in (c).\\
\\
{\bf Fig.3}  Behaviour of occupation probabilities $|\langle \psi(t) | n \rangle|^2$ of Floquet eigenstates at time $t=0$ (upper panels) and over one oscillation cycle (maps in the lower panels) in the ac-driven lattice of Fig.1(a) corresponding to (a) a BOC mode, (b) a resonance state, and (c-e) the three BIC states labelled by letters A,B and C in Fig.2 . Parameter values are: (a,b) $\kappa / \omega=2$, $\Gamma=2.28$;  (c) $\kappa / \omega=2$, $\Gamma=1.9940$;  (d) $\kappa / \omega=1$, $\Gamma=2.5195$;  (e) $\kappa / \omega=0.3$, $\Gamma=2.3800$. The insets in the upper panels (c-e) show an enlargement of the tails of the BIC modes, corresponding to $D=h_2 / h_1 <10^{-6}$.\\
\\
{\bf Fig.4} 
Numerically-computed evolution of a Gaussian wave packet (snapshot of $|\langle \psi(t) | n \rangle |^2$) in the ac-driven lattice of Fig.1(a) for parameter values (a) $\kappa/ \omega=0.3$, $\Gamma=2.308$, and (b) $\kappa / \omega=2$, $\Gamma=1.994$. The initial condition is $\langle   n | \psi(0) \rangle  \propto \exp[-(n-n_0)^2/w_0^2-i \pi n /2]$ with $w_0=4$ and $n_0=-20$.\\

\newpage

\includegraphics[width=12cm]{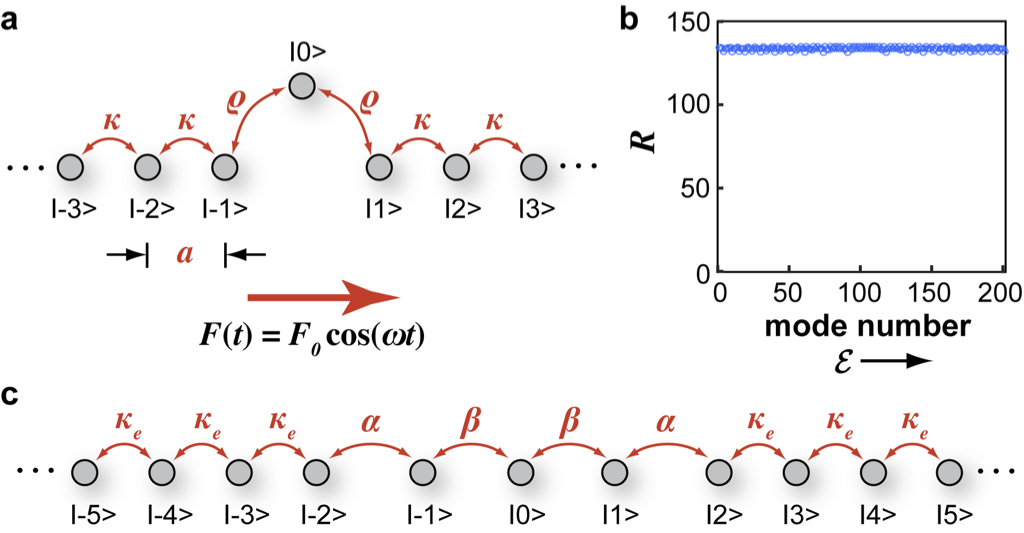}

\newpage

\includegraphics[width=18cm]{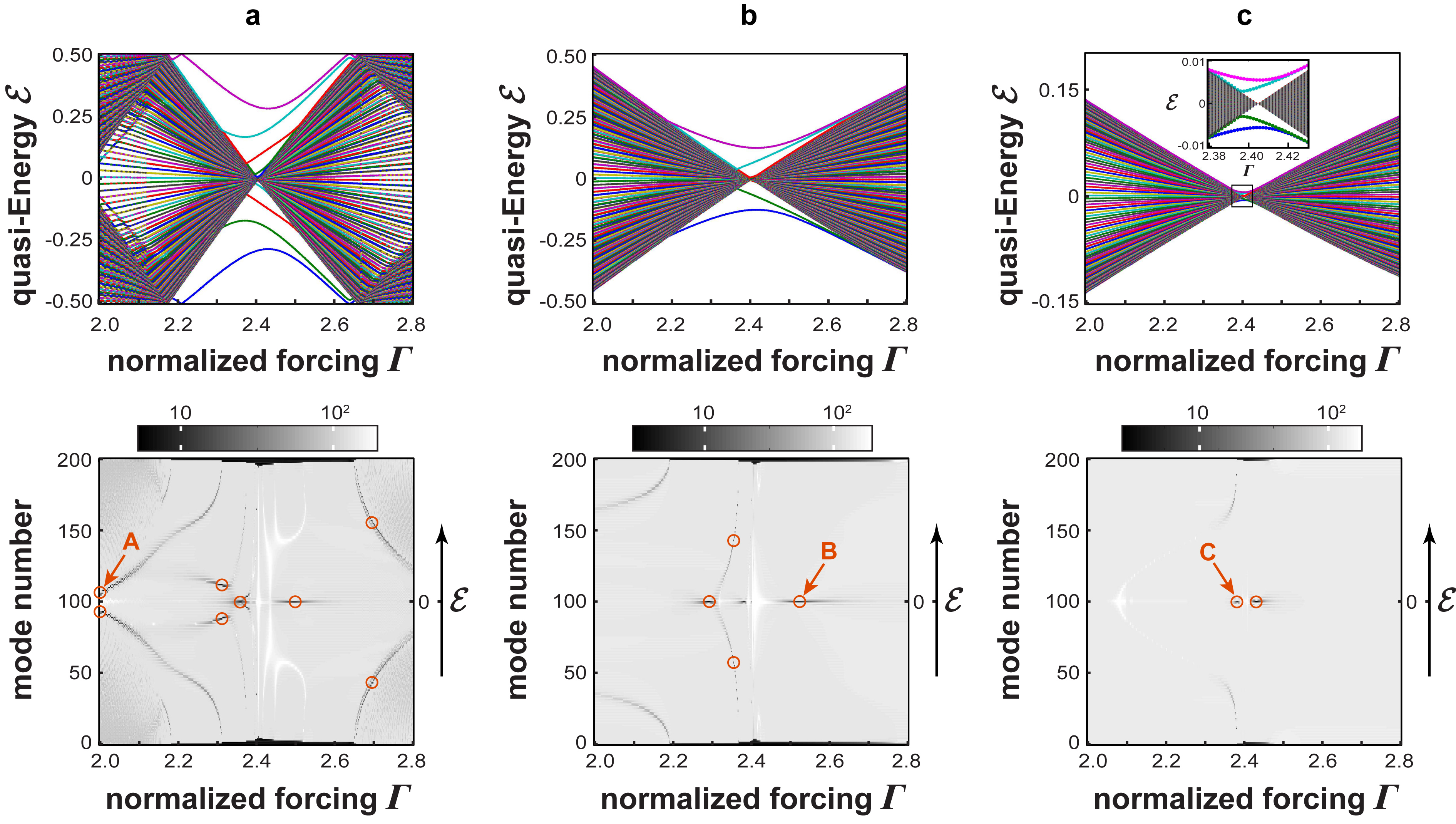}

\newpage

\includegraphics[width=18cm]{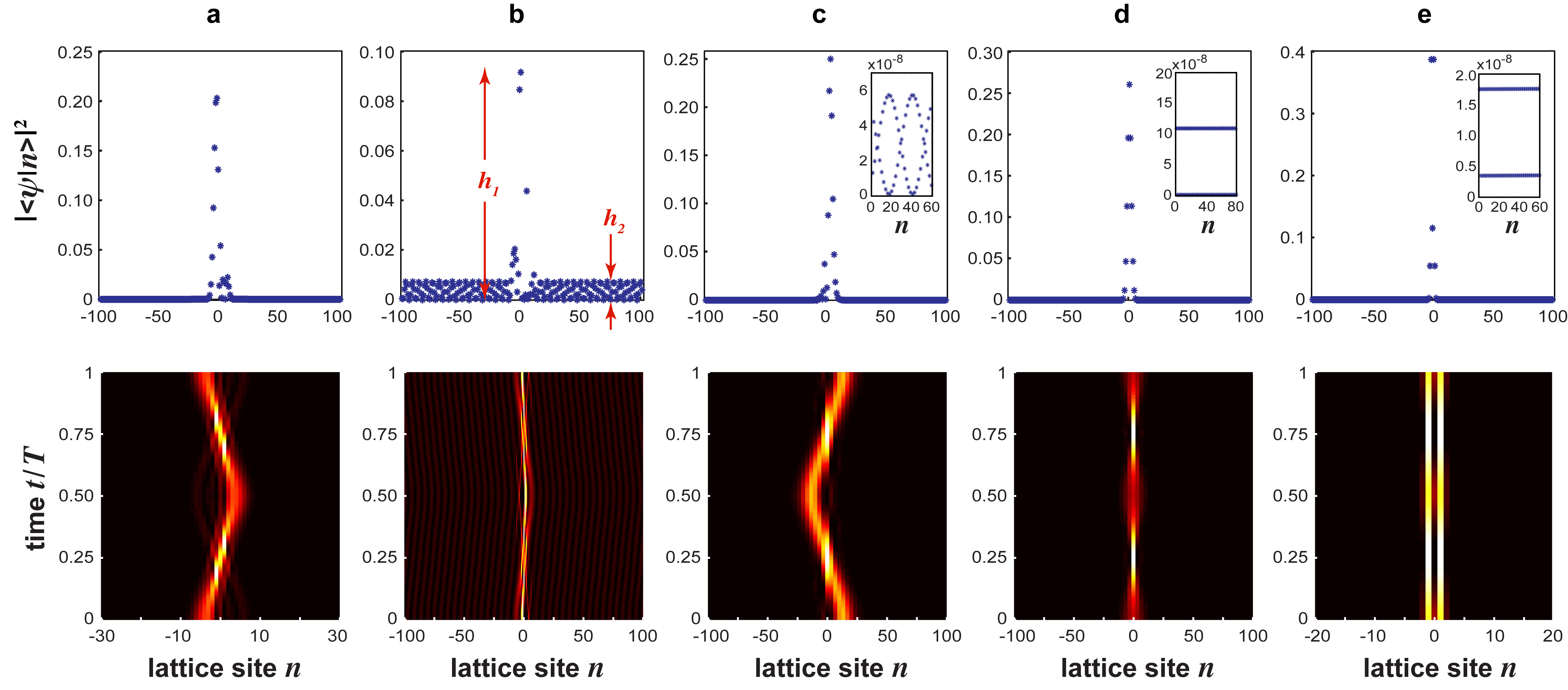}

\newpage

\includegraphics[width=16cm]{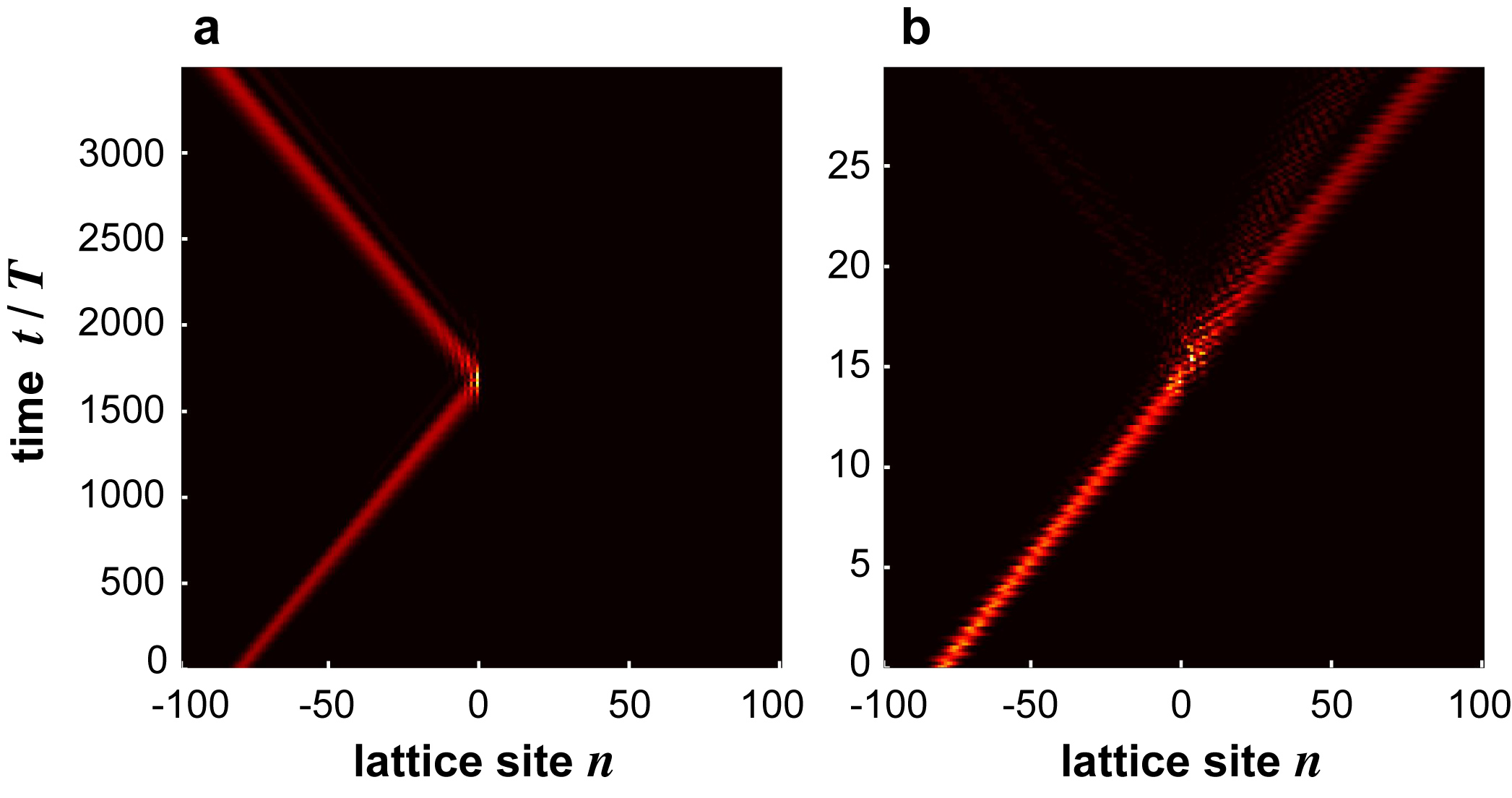}

%\begin{figure*}
%\includegraphics[width=12cm]{Fig1}
%\caption{ (a) Schematic of an ac-driven tight-binding lattice with inhomogeneous hopping rate $\rho< \kappa$ between lattice sites $|0 \rangle$ and $| \pm 1 \rangle$. (b)
%Participation ratio $R$ of the eigenstates of the undriven lattice comprising $N=201$ sites for $\rho / \kappa=0.7$. The mode number on the horizontal axis is ordered for increasing values of the energy $\mathcal{E}$, from $\mathcal{E}=-2 \kappa$ (lower axis limit) to $\mathcal{E}=2 \kappa$ (upper axis limit). (c) Effective static lattice model in the high-frequency regime. The effective hopping rates $\kappa_e$, $\alpha$ and $\beta$ are given in the text by Eq.(3).}
%\end{figure*}

\end{document}